\documentclass[sigconf,screen,nonacm]{acmart}

\AtBeginDocument{%
  }

\usepackage{xcolor}
\usepackage{url}

\setcopyright{acmlicensed}
\copyrightyear{2018}
\acmYear{2018}
\acmDOI{XXXXXXX.XXXXXXX}
\acmConference[Conference acronym 'XX]{Make sure to enter the correct
  conference title from your rights confirmation email}{June 03--05,
  2018}{Woodstock, NY}
\acmISBN{978-1-4503-XXXX-X/2018/06}

\begin{document}

\title{On the Surprising Efficacy of LLMs for Penetration-Testing}

\author{Andreas Happe}
\email{andreas.happe@tuwien.ac.at}
\orcid{0009-0000-2484-0109}
\affiliation{%
  \institution{TU Wien}
  \city{Vienna}
  \country{Austria}
}

\author{Jürgen Cito}
\email{juergen.cito@tuwien.ac.at}
\orcid{0000-0001-8619-1271}
\affiliation{%
  \institution{TU Wien}
  \city{Vienna}
  \country{Austria}
}

%\author{
%\IEEEauthorblockN{Andreas Happe}
%\IEEEauthorblockA{\textit{TU Wien} \\
%Vienna, Austria\\
%andreas.happe@tuwien.ac.at}
%\and
%\IEEEauthorblockN{Jürgen Cito}
%\IEEEauthorblockA{\textit{TU Wien} \\
%Vienna, Austria\\
%juergen.cito@tuwien.ac.at}
%}

\begin{abstract}
This paper presents a critical examination of the surprising efficacy of Large Language Models (LLMs) in penetration testing. The paper thoroughly reviews the evolution of LLMs and their rapidly expanding capabilities which render them increasingly suitable for complex penetration testing operations. It systematically details the historical adoption of LLMs in both academic research and industry, showcasing their application across various offensive security tasks and covering broader phases of the cyber kill chain. Crucially, the analysis also extends to the observed adoption of LLMs by malicious actors, underscoring the inherent dual-use challenge of this technology within the security landscape.

The unexpected effectiveness of LLMs in this context is elucidated by several key factors: the strong alignment between penetration testing's reliance on pattern-matching and LLMs' core strengths, their inherent capacity to manage uncertainty in dynamic environments, and cost-effective access to competent pre-trained models through LLM providers.

The current landscape of LLM-aided penetration testing is categorized into interactive 'vibe-hacking' and the emergence of fully autonomous systems. The paper identifies and discusses significant obstacles impeding wider adoption and safe deployment. These include critical issues concerning model reliability and stability, paramount safety and security concerns, substantial monetary and ecological costs, implications for privacy and digital sovereignty, complex questions of accountability, and profound ethical dilemmas. This comprehensive review and analysis provides a foundation for discussion on future research directions and the development of robust safeguards at the intersection of AI and security.
\end{abstract}



%% Keywords. The author(s) should pick words that accurately describe
%% the work being presented. Separate the keywords with commas.

\maketitle

\section{Introduction}

Upholding an organization's IT security has been problematic since the rise of enterprise networks in the early 2000s. One of the most publicly visible type of security incidents are ransomware attacks. Current estimates indicate that 72\% of business were affected by ransomware between 2018--2023~\cite{statista_ransomeware}. Related losses are estimated to reach \$57 billion for 2025 or \$6.5 million per hour~\cite{Ransomware2}. The trend indicates a worsening of this situation with estimates of ransomware incidents happening every 2 seconds in 2031, up from 11 seconds in 2021~\cite{Ransomware2}.

A common way of preventing security incidents is to test one's own defenses through penetration-testing. The capability of performing sufficient penetration-testing is severely limited by the amount of available offensive personnel, i.e. ISC2 estimates that currently 4.7 million cybersecurity experts are missing from the workforce. The year-over-year change indicates that this gap is still increasing, e.g., by 19.1\%~\cite{isc2_workforce} in 2025, this indicates a massive need to make existing security penetration-testers more effective, or even to automate time-consuming parts of penetration-testing, reducing the need for manual work by human specialists. As will be shown in Section~\ref{history}, Large Language Models (LLMs) have been increasingly investigated by both industry and academia to fulfill this need.

In this paper, we reflect on the first two years (2023--2025) of LLM-aided penetration-testing. We use the Background Section (Section~\ref{background}) to create a common understanding of penetration-testing and highlight important milestones within the evolution of LLMs. In Section~\ref{history}, we give a short history of how LLMs were used for penetration-testing, separated into academic research and adoption by industry. After we have shown their clear and present interest in LLMs, we discuss why LLMs are a good fit for penetration-testing and summarize the current status-quo (Sections~\ref{fit} and ~\ref{status_quo}). Subsequently, we highlight obstacles that prevent further adoption of LLMs before we conclude with a final section highlighting potential remediation means, also known as research opportunities, for the mentioned obstacles.

\section{Background}
\label{background}

We introduce penetration-testing, show how pre-LLM AI systems were used by adversaries, and conclude with a high-level overview of the evolution of LLMs.

\subsection{Penetration-Testing}
\label{pentesting_pattern_matching}

During penetration-testing highly-skilled professionals try to break systems to uncover vulnerabilities so that those can be mitigated by defensive personnel before malicious actors can exploit them. We focus on the practitioners' daily work~\cite{Happe_2023}, not on security researchers that seek to find new vulnerabilities and attack vectors that practitioners will subsequently exploit.

Enterprise networks are typically comprised of Microsoft Active Directory, with industry studies indicating that over 90\% of Global Fortune 1000 companies are using it as their primary means of authenticating and authorizing users~\cite{ad}. Having this common technology stack, or mono-culture, for enterprise network architectures allows penetration-testers to apply knowledge learned during prior penetration-tests to new assignments by matching common problems and insecure configurations. Practitioners also describe that they can apply knowledge learned during educational attacking of simulated systems (CTFs) to real-world systems~\cite{Happe_2023}, further enforcing the idea that \textbf{penetration-testing is often based on pattern-matching}.

\subsection{Pre-LLM Usage of AI for Offensive Security}

Interest into applying machine-learning techniques to offensive security tasks precedes the rise of LLMs. Mirsky et al.~\cite{mirsky2023threat} performed a literature review and surveyed experts from academia, industry, and government on the potential threat of offensive AI to organizations. The initial version of the paper was released to arXiv in July 2021, a revised version was published in January 2023. As ChatGPT went public in late November 2022, this publication describes the state-of-the-art pre-LLM.

The authors identify 33 AI offensive capabilities and group them into \textit{Automation}, \textit{Campaign Resilience}, \textit{Credential Theft}, \textit{Exploit Development}, \textit{Information Gathering}, and \textit{Stealth}. Of these categories, the top 3 identified capabilities were \textit{Exploit Development}, \textit{Social Engineering}, and \textit{Information Gathering}. Their survey indicated that, overall, academia and industry felt that impersonation (and thus social engineering) was the biggest threat, mirroring another user study of professional penetration-testers~\cite{Happe_2023} that highlighted the potential of machine-learning for phishing, a subtask of social engineering.

They conclude that offensive AI primarily impacts the initial steps of the cyber kill chain, focusing on reconnaissance, resource development, and initial access as ``\textbf{AI technologies are not mature enough to create agents able to carry on attacks that proceed without human supervision and aid}''. In 2021, their outlook on the near future was that ``\textbf{we aren’t likely to see botnets that can autonomously and dynamically interact with a diverse set of complex systems (like an organization’s network) in the near future}''.

We will show in this paper that the rise of LLMs has accelerated the adoption of machine-learning for offensive security and has fulfilled some of the paper's predictions. While we focus upon penetration-testing (roughly comprising the \textit{automation} and \textit{campaign resilience} categories of their study), our section on malicious industrial use of LLMs (Section~\ref{industrial:blackhats}) will show that their predictions about social engineering, exploit development, and information gathering became reality. Furthermore, while not employed by industry yet, LLMs enabled academic prototypes (Section~\ref{academic_research}) to cover more phases of the cyber kill chain, including privilege escalation and lateral movement. Dwelling deeper on their list of identified offensive AI capabilities, we see the following \textit{Automation} capabilities well-covered by research: \textit{Attack Adaption}, \textit{Attack Coordination}, \textit{Next Hop Targeting}, \textit{Point of Entry Detection}. Of \textit{Campaign Resilience}'s capabilities, we see \textit{Campaign Planning} covered; of the other techniques some are implicitly covered by the prototypes highlighted in Section~\ref{academic_research}, e.g., \textit{Virtualization Detection} and \textit{Password Guessing}.

\subsection{Evolution of LLMs}
\label{evoluation_of_llms}

In November 2022, OpenAI made ChatGPT publicly available~\cite{intro_chatgpt} and thus sparkled public interest into LLMs. Their API allowed integration of LLMs into existing tools and workflows.

Over time, LLMs gained advanced capabilities~\cite{khan2025advances} such as \textit{tool-calling} or \textit{structured-output}. The former allows LLMs to call user-supplied functions, typically used to interact with their environment. Structured output allows easier integration of LLM's output into the caller's system. Both of them have lead to \textit{agentic AI}, systems in which agents issue commands to interact with their environment to solve their given tasks, often autonomously. \textit{Retrieval Augmented Generation} (RAG) and using the ever-increasing LLM context size for \textit{In-Context Learning} allowed LLMs to integrate additional background knowledge~\cite{lee2024longcontextlanguagemodelssubsume} without having to expensively retrain models. It became feasible to operate smaller models on commodity hardware on-edge~\cite{touvron2023llama}. Prompt-Engineering used techniques such as \textit{Chain-of-Thought}~\cite{wei2022chain} (CoT) to improve the efficacy of LLMs in complex tasks.

Another big step forward was the introduction of reasoning LLMs, e.g., by OpenAI introducing the o1 model in 2024~\cite{intro_o1}. Reasoning models are pre-trained to incorporate CoT during inference, reducing the dependence on prior prompt-engineering techniques for solving complex tasks. Recent contested research~\cite{petrov2025proof,shojaee2025illusion} indicates that reasoning models do not perform reasoning similar to humans, but that LLMs are ``merely'' \textbf{getting even better at pattern-matching} than they were before. Concurrent research indicates that reasoning LLMs can introduce problems with \textit{tool-calling} and \textit{instruction-following}~\cite{li2025thinking}.

Latest advancements further improve the ability of LLMs to interact with their environment or with each other. The \textit{Model Context Protocol} (MCP), originally proposed by Anthropic in 2024~\cite{intro_mcp}, standardized integration of tools into LLMs, leading to an explosion of available tool-integrations. Recent interest into \textit{Multi-Agent Systems}, e.g. Google's A2A~\cite{intro_a2a}, uses multiple collaborating LLMs to solve complex tasks~\cite{kong2025surveyllmdrivenaiagent}.

\section{A Short History of Using LLMs for Penetration-Testing}
\label{history}

Give the rapid evolution for LLMs, we want to highlight their impact on security with a focus on penetration-testing. We differentiate between academic research and industry adoption.

For academic research, we used Google Scholar to identify survey papers on offensive use of LLMs~\cite{hassanin2024comprehensiveoverviewlargelanguage, YAO2024100211, jin2024llmsllmbasedagentssoftware, 10613562, xu2024largelanguagemodelscyber, motlagh2024largelanguagemodelscybersecurity, dube2024large, 10.1145/3659677.3659749, yigit2024reviewgenerativeaimethods, zhang2024llmsmeetcybersecuritysystematic, 10992500}. We included papers from these surveys if they were using LLMs to perform penetration-testing, and analyzed them in chronological order for their novelty. The analysis of industrial use of LLMs is based on articles posted on security-specific news-sites\footnote{\url{https://www.bleepingcomputer.com/} and \url{https://www.darkreading.com/}} mentioning the use of LLMs, as well as on abuse reports provided by LLM providers.

\subsection{Academic Research}
\label{academic_research}

We group academic research based on the initial publication year of the respective publication, typically using the date of their initial upload to arXiv.

\subsubsection{Initial Forays (2023)}

Given our described methodology, the first paper that used LLMs for penetration-testing was \textit{Getting pwn'd by AI} by Happe et al.~\cite{happe2023getting} in July 2023. They differentiated between two use-cases. On a strategic level, they used chatGPT to devise an attack plan and gather information about a target organization. On an operational level, they introduced an autonomous penetration-testing prototype capable of performing privilege-escalation attacks against a vulnerable Linux virtual machine.

Deng et al.~\cite{deng2024pentestgpt} published \textit{pentestGPT} in August 2023. Their prototype integrates human operators with an LLM to interactively hack CTF boxes. They were the first paper to explicitly state that LLMs have sufficient inherent capabilities for hacking but have problems with context-management, manifesting in missing long-term memory, recency bias, and hallucinations. They propose the Pentest-Task-Tree, a hierarchical todo list, to alleviate these problems.

In October 2023, Happe et al.~\cite{happe2024llms} published \textit{LLM as Hackers}. While still focusing on Linux privilege-escalation attacks, they exchanged the single vulnerable VM with a benchmark comprised of different privilege-escalation vulnerability classes to further analyze the capabilities of LLMs. They focus on context-management, summarization capabilities, in-context learning for providing background knowledge, the impact of high-level guidance, and included small language models in their evaluation. Their results indicate that small language models were not feasible for penetration-testing, highlight LLMs' problems with complex multi-step attacks, and the importance of high-level strategy/guidance mechanisms for the overall performance of LLMs.

\subsubsection{Broadening Domains and Embracing Newer LLM Capabilities (2024)}

In 2024, LLM-capabilities improved through the introduction of function-calling and structured-output. Supported context sizes increased from typical 4--16 kTokens in 2023 to typically 64--128 kTokens, with some models allowing for even larger context sizes, e.g., Google's Gemini-1.5 model allowed for a context size of up to one million tokens in February 2024.

In February 2024, Fang et al.~\cite{fang2024llmagentsautonomouslyhack} published \textit{LLM Agents can Autonomously Hack Websites}, extending targets to web sites and showing that LLMs were able to autonomously find vulnerabilities within them. They incorporated both in-context learning and were the first to use function-calling in their prototype. Also in February, Shao et al.~\cite{shao2024empiricalevaluationllmssolving} published \textit{An Empirical Evaluation of LLMs for Solving Offensive Security Challenges}. They analyze the capacity of LLMs for solving CTF challenges in interactive and autonomous settings. Similar to Fang et al., their prototype incorporated function-calling. They conclude that LLM-driven prototypes produced similar results compared to human penetration-testers.

In March, Xu et al.~\cite{xu2024autoattacker} introduced \textit{AutoAttacker} focusing on post-breach attacks using the Metasploit attack framework. They explicitly mention the use of RAG for background knowledge storage. They are also the only paper within our analysis that explicitly mentioned the need for jailbreaks, i.e., being caught by LLM provider's safety filters and needed to bypass these filters by utilizing roleplay prompting~\cite{johnson2024generation}.

Fang et al. analyzed the capability of LLMs for exploit development, i.e., their capability to create exploits for both one-day\footnote{A zero-day vulnerability is unknown to the vendor, and thus there is no patch, mitigation, or fix available to address it. One-day vulnerabilities are known vulnerabilities for which a patch or mitigation is available but hasn’t yet been applied.}~\cite{fang2024llmagentsautonomouslyexploit} and zero-day vulnerabilities~\cite{fang2024teamsllmagentsexploit}. In their initial paper, they used a GPT-4 based ReAct-agent and highlighted the need for better planning and improved exploration capabilities. In their latter paper, they implemented a hierarchical planning system using multiple task-specific agents that were provided background knowledge through in-context learning. Their results indicate that hierarchical planning improved penetration-testing results by a factor of 6, while task-specific agents and in-context learning both doubled the performance.

Finally, in October 2024, Gioacchini et al.~\cite{gioacchini2024autopenbenchbenchmarkinggenerativeagents} introduced \textit{AutoPenBench}. They use function-calling and structured-output to autonomously solve CTF challenges. Their results indicate that LLMs were able to solve challenges if similar tasks were well documented publicly through blog posts and walk-throughs.

\subsubsection{Breaching Out (2025)}
\label{history:breaching}

In January 2025, Kong et al.~\cite{kong2025vulnbot} introduced \textit{VulnBot}, a multi-agent autonomous prototype using a penetration-testing task graph as internal storage mechanism for creating high-level strategies. They also incorporated RAG for providing background knowledge to the agent. Also in February, Singer et al.~\cite{singer2025feasibility} published \textit{On the Feasibility of Using LLMs to Execute Multistage Network Attacks}, switching from the single-host attacks performed by previous papers to attacking complex multi-stage networks. They introduce a tool abstraction layer that simplifies tool usage for LLMs, indicating that this abstraction enables smaller LLMs to successfully perform penetration-testing, while improving the hacking results of larger models.

In February 2025, Happe et al.~\cite{happe2025llmshackenterprisenetworks} published \textit{Can LLMs Hack Enterprise Networks?}, replacing single targets with a real-life Microsoft Active Directory enterprise network. Their prototype consists of a high-level strategy component using a penetration task tree, and a low-level ReAct-based task execution agent. They use tool-calling and structured-output and analyze the capabilities offered by reasoning LLMs. Their results indicate that modern models contain enough penetration-testing knowledge to perform autonomous hacking without providing background-knowledge through RAG. Their results indicate that LLMs have sufficient hacking capabilities but that results lack consistency, i.e., vary between testruns. They highlight models' auto-repair capabilities and conclude that the costs of using LLMs for penetration-testing compares favorable to human penetration-testers.

\subsubsection{Specialized LLMs for Security Tasks} In parallel, LLMs were fine-tuned for security-tasks~\cite{pratama2024cipher,whiterabbitneo}. As their makers typically do not publish these models, or, if published, they lack capabilities such as tool-calling, these specialized models were not used within the reviewed publications.

\subsection{Industry Adoption}
\label{history:industry}

We differentiate between white-hats trying to improve security, and black-hats trying to use LLMs to exploit security vulnerabilities.

\subsubsection{Benign White-Hats}
\label{industrial:whitehats}

There has been interest in using LLMs to either accelerate tedious tasks during vulnerability research, to increase test coverage within analyzed projects, and to cover more projects with vulnerability research.

Google operates OSS-Fuzz\footnote{\url{https://github.com/google/oss-fuzz}} which provides continuous fuzzing for open-source projects. In order to fuzz a project, complex fuzzing harnesses have to be created. OSS-Fuzz started to use AI for creating and testing these fuzz harnesses using AI~\cite{ossfuzz_harness} in August 2023, and reported 26 vulnerabilities detected with help of AI in November 2024~\cite{ossfuzz_harness2}. Their blog post highlights how LLMs were used to create and debug fuzz harnesses, leading to increased fuzzing coverage. In addition, LLMs were used to analyze the traces gathered by the fuzzing process.

We will use the cooperation between Google's Project Zero, a team of security-analysts tasked with finding zero days, and Google DeepMind as an example study for using LLM agents in security research. In June 2024, Project Zero detailed \textit{Project Naptime}~\cite{naptime}, a LLM-powered vulnerability research framework. They utilized Chain-of-Thought, an interactive environment, specialized tools for debugging, and provide an external verification environment to validate found vulnerabilities. To explore multiple vulnerability hypotheses, instead of implementing a high-level strategizing loop, they advocate for a sampling strategy that explores multiple hypotheses through independent trajectories. In November 2024, they were able to disclose the first real-world vulnerability found through their agent, now called \textit{Big Sleep}: an exploitable stack buffer underflow in SQLite~\cite{bigsleep}.

LLM-use is not limited to large-scale companies such as Google. Sean Heelan used OpenAI's o3 model to find CVE-2025-37899~\cite{smb}, a remote zero-day in the Linux kernel's SMB implementation (a commonly used network file-system). They provided only a subset of the SMB code to o3 as including the full kernel code would exceed o3's context size. They instructed the LLM to specifically search for use-after-free vulnerabilities~\cite{xu2015collision}, gave a high-level overview of the SMB module, and provided a threat model. They then ran the resulting prompt 100 times, resulting in 8 trajectories correctly identifying the vulnerability, indicating that LLMs have sufficient capabilities for finding zero-days, but lack reliability.

LLMs are also used for non-exploitation purposes. Matt Adams released \textit{StrideGPT}\footnote{\url{https://github.com/mrwadams/stride-gpt}}, a LLM-powered automated threat-modeling tool. Daniel Miessler, a well-known security professional, provides \textit{fabric}\footnote{\url{https://github.com/danielmiessler/fabric}}, an open-source framework for augmenting humans with AI. In his opinion, ``\textit{AI doesn't have a capabilities problem---it has an integration problem}''.

This list does not include indirect industry use, i.e., security professionals using LLMs instead of dedicated search systems for security-specific information-retrieval. Anecdotally, LLM systems are also used during reporting.

\subsubsection{Malicious Black-Hats}
\label{industrial:blackhats}

Academic research indicated early uptake of LLMs by malicious actors, often facilitated within the Darknet~\cite{lin2024malla, firdhous2023wormgpt}. Specialized LLMs employing neither ethical guardrails nor safety filters were offered to help with exploit development, social engineering, and information gathering. Monitoring this malicious use of LLMs is complicated by their provider's inherent covert and illegal operation.

Public cloud-backed LLM providers have started to periodically publish abuse reports. We analyze malicious tasks included in the reports provided by OpenAI~\cite{openai_abusereport_feb_2024, openai_abusereport_october, openai_abusereport_feb, openai_abusereport_june}, Anthropic~\cite{anthropic_abuse}, and Google~\cite{google_abuse}. Overall, they show a similar theme: threat actors use LLMs to accelerate and optimize their work, but they do not use them to create novel methods of attack. Threat actors use LLMs for information gathering similar to using search engines, employ them for developing and debugging malicious software, and use them to generate content for social engineering and phishing attacks.

Presumed state-level actors use LLMs for covert \textit{Influence Operations} (IO) trying to perform election tinkering, sway public opinion especially in and around conflict zones, discredit political activists and parties, attack independent media, sow discontent within populations, and polarize existing population sub-groups. They use LLMs to rewrite articles from genuine news sources with a particular political perspective or tone. Anthropic highlighted an \textit{Influence-as-a-Service} operation in their March 2025 report~\cite{anthropic_abuse, anthropic_influence}, detailing a semi-autonomous LLM-driven system that used approximately 100 fake social-media puppet accounts to influence opinion. AI was used to make both strategic and tactical decisions when and how to employ these social accounts and incorporated LLM-powered image generators.

Advanced Persistent Threats (APTs) use LLMs to aid development of malware and backdoors, analyze defensive capabilities and perform deceptive employment schemes.\footnote{These are a form of social engineering attack in which the attacker applies for a job interview to gain access to the target organization.} As Anthropic states~\cite{anthropic_abuse}, LLMs ``\textit{flatten the learning curve for malicious actors}''. OpenAI' highlighted in its June 2025 report~\cite{openai_abusereport_june} that threat actors start to research into LLM-driven penetration-testing.

\section{On the Surprising Efficacy of LLMs for Penetration-Testing}
\label{fit}

In this section we speculate why LLMs have become a part of the vanguard for automated penetration-testing.

There are few empirical studies on the work practices of penetration-testers and their decision making processes~\cite{Happe_2023}. Thus, we include our own experiences (one of the authors has been a professional penetration-tester for 13 years and taught penetration-testing both in academic and industrial settings). We encourage further empirical research into hackers' work and expect that future findings will support our hypotheses.

\subsection{Penetration-Testing Resembles Pattern-Matching}

We focus on security practitioners within this work: they are the professionals that perform daily penetration-tests to find vulnerabilities in enterprise networks and web-applications.

\subsubsection{Pattern-Matching is a Substantial Part of Penetration-Testing}

Empirical studies with security professionals~\cite{Happe_2023} list examples of practitioners applying pattern-matching, e.g., identifying vulnerable areas or operations through knowledge gained from CTF exercises, applying knowledge learned as software developer, using knowledge from prior interactions with the customer and their systems, and using vulnerabilities that the penetration-tester has exploited before.

One example of pattern-matching in penetration testing are \textit{Vulnerability Assessments}. During those, software version numbers, detected from service banners or application errors messages, are compared against well-known vulnerabilities. If a match occurred, a potentially available exploit is prepared, i.e., its options are filled with data gathered from the target environment, and execute to exploit the expected vulnerability. Pattern matching occurs at multiple levels: detecting error messages, matching them to the vulnerability catalog, matching the right configuration options, and matching the exploit's output to the expected behavior.

Another low-level occurrence of pattern-matching happens during analyzing of web-application responses containing error messages. For example, LLMs can match error messages that indicate database issues (SQL injection attacks) to knowledge in their training data, and subsequently successfully exploit these vulnerabilities~\cite{fang2024llmagentsautonomouslyhack}.

On a higher level, creating an attack-strategy is also grounded in patterns seen during security testing. Interviews with professional penetration-testers indicate that they encounter similar insecure configurations during assignments, create a hypothesis about the network's security, and select attacks based upon that~\cite{Happe_2023}. For example, if penetration-testers encounter unsigned NTLMv2-Hashes during an enterprise network security test, they assume that the overall target network matches security bad-practices from 10 years ago and attempt matching attacks such as pass-the-hash attacks.

\subsubsection{The Target System Landscape is Homogeneous} Pattern-matching is only feasible if penetration-testing targets exhibit similarities---otherwise there would be insufficient features to base the pattern-matching on. Fortunately, when looking at enterprise networks, over 90\% of Global Fortune 1000 companies use the same underlying technology stack (Section~\ref{pentesting_pattern_matching}). Security practitioners note that they encounter the same security vulnerabilities and insecure configurations during assignments~\cite{Happe_2023}. When looking at web penetration-testing, the landscape is more diverse. Applications utilizing the same technology stack, e.g., the same web framework, are similar implementation-wise and often exhibit similar vulnerabilities. Architectures between different technology stacks also show similarity through common architecture design patterns.

\subsubsection{LLMs exceed in Pattern-Matching}

One side-effect of the ongoing discussion about LLMs' reasoning capabilities is that researchers agree that LLMs exceed at pattern-matching~\cite{shojaee2025illusion,mirchandani2023largelanguagemodelsgeneral,schindler2025llmbaseddesignpatterndetection}. Given the previous section, this implies that they are well-suited for penetration-testing.

Research implies that LLMs are able to solve tasks if examples in their training data resemble those tasks~\cite{gioacchini2024autopenbenchbenchmarkinggenerativeagents}. Given the described semi-monocultures, target IT environments should resemble each other. Furthermore, there is ample publicly available penetration-testing background knowledge available from technical blog posts, incident reports, and CTF walk-throughs and thus included in common training data sets. We further note that common walkthrough formats, i.e., providing step-by-step instructions including reasoning steps and tool usage examples, structurally resembles trajectories used to train LLMs and thus are well-suited to train LLMs.

\subsection{LLMs Inherently Cope With Uncertainty}

\subsubsection{Uncertainty During Penetration-Testing}

Security practitioners routinely deal with uncertainty~\cite{Happe_2023, hu2024uncertainty}. Examples given for sources of uncertainty include interpreting misleading tool outputs and target system responses, negative side-effects like target systems becoming unresponsive due to exploits, incomplete information about the target environment, and invalid but not falsified assumptions about the target system's behavior and security.

\subsubsection{LLMs' Pattern-Matching Copes with Uncertainty}

Pattern-matching is inherently capable of dealing with uncertainty. During a penetration-test, a LLM-driven hacking prototype will sample the target environment and create a representation of its view of the target world, typically as a text-based representation. This world view is typically included in subsequent LLM invocations. Compared to deterministic rule-based systems, LLMs are able to ignore parts of their world-view through the pattern-matching process. This is beneficial during penetration-testing in realistic network scenarios where actions influence the state of the network, e.g., where a failed attack might lead to locked accounts or crashed network servers, changing the ground truth. While traditional systems need to manually invalidate their world view, LLMs implicitly do this through their pattern-matching approach.

\subsection{The Costs of Using LLMs}

When adopting machine learning, there's always the question of costs for creating and operating the system, as well as for creating a training data-set and utilizing it for training.

\subsubsection{LLM-Makers are Front-Loading Creation Costs}

As shown through-out our review of academic research (Section~\ref{academic_research}), off-the-shelf LLMs already contain sufficient background knowledge to perform penetration-testing, alleviating the need to costly train new security-specific LLMs. Even if specific additional knowledge is needed, In-Context Learning and RAG (Section~\ref{evoluation_of_llms}) offer cost-effective alternatives to training a model from scratch.

A model alone does not make a penetration-testing prototype---integration with its environment is also required. The LLM ecosystem provides easy access to development libraries/frameworks. Technologies, such as function-calling or MCP, make integration efficient from a development perspective.

\subsubsection{LLM-Providers Enable Cost-Effective Inference}

Running LLMs using cloud-based LLM-providers is cost efficient, e.g., Happe et al.~\cite{happe2025llmshackenterprisenetworks} listed operational costs for their penetration-testing prototype running from \$0.10 to \$11.64 depending on the used LLM.

Given the sensitive nature of using LLMs for hacking, LLM providers imposing stricter safety guards would negate this benefit. Of the reviewed academic publications, a single paper mentioned problems with safe-guards~\cite{xu2024autoattacker}---and those were easily bypassed by simple techniques such as roleplay-prompting~\cite{xu2024autoattacker,johnson2024generation}.

\subsubsection{Costs of Running to Stand Still}

In evolutionary biology, the Red Queen's hypothesis proposes that species must continuously adopt and evolve to survive while pitted against ever-evolving opposing species~\cite{van1973new}. In penetration-testing, we have active adversaries (defensive blue teams) that adapt to our attacks and evolve their defenses based on our activities, e.g., develop new intrusion-detection (IDS/IPS) or endpoint detection and response (EDR) capabilities.

Security tooling, esp. in case of covert C2 frameworks or vulnerability scanners, impose a high maintenance cost as they have to be adopted to new vulnerabilities, current trends, and evolved adversary measures. Penetration-testers also have to continuously improve, e.g., learn new attack vectors or how to circumvent novel counter-measures. Delegating parts of these costs to the LLM-maker, as newer training data will inherently incorporate these new techniques, is thus tempting to time-poor penetration-testers.

\subsection{Additional Beneficial Capabilities}

We want to high-light additional LLM capabilities that, while not required for successful penetration-testing, are beneficial.

\subsubsection{Inter-Context Attacks} Compared to traditional tooling, LLMs provide multi-modal or inter-context capabilities~\cite{happe2025llmshackenterprisenetworks}. For example, if a LLM detects a web-server during an enterprise network penetration-test, it will switch to a web-testing context and perform a web penetration-test. They are able to detect passwords in text files~\cite{happe2025llmshackenterprisenetworks} and utilize them during subsequent penetration-testing steps. This is a time-consuming task, typically performed by human penetration-testers~\cite{Happe_2023}.

\subsubsection{Hallucinations often not deemed catastrophic} LLMs are prone to hallucinations, e.g., they invent untrue facts. In small amounts, this can be beneficial during penetration-tests as it is similar to human penetration-testers trying out hypotheses about the security of their target systems. Thus, limited hallucinations are not as problematic for the penetration-testing use-case compared to other software-engineering use-cases.

\section{Status-Quo: Vibe-Hacking and Autonomous Agents}
\label{status_quo}

We want to highlight the current status-quo of LLM-driven penetration-testing. We structure this into interactive use of LLMs (\textit{vibe-hacking}) and in prototypes that use LLMs to autonomously hack systems.

\subsection{Vibe-Hacking}

As Section~\ref{history:industry} has shown, industrial interactive uptake of LLMs is already occurring. Borrowing from \textit{vibe-coding}, this interactive delegating of tedious tasks has become known as \textit{vibe-hacking}. Levels of autonomy are diverse, ranging from chat-based LLMs for information-retrieval and exploit-code generation, over co-pilot inspired augmenting agents, to systems influenced by pentestGPT~\cite{deng2024pentestgpt} where humans are responsible for oversight. CTF challenges are often created by the same authors, exhibiting patterns in their structure. Anecdotally, CTF players use OpenAI's custom GPTs support to create LLMs trained with previous challenges created by well-known authors and use that knowledge for future challenges. Compared to other automation approaches, currently vibe-hacking keeps the human in the loop and thus incorporating an important safety feature.

While \textit{vibe-hacking} is not often discussed online, \textit{vibe-coding} is, current opinions range from AI Angst to AI enthusiasm~\cite{aiangst}. The former highlights the potential for developer burnout, catastrophic failures, ethical problems, and increased maintenance costs~\cite{noai}; while the latter highlight developer efficiency gains and reduction of tedious work. While our use-case, penetration-testing, does not include high-maintenance overheads, safety concerns are paramount~\cite{ainuts}.

We foresee that, over time, more and more complex tasks will be delegated to LLM agents, culminating in LLMs autonomously performing complex multi-step tasks. This is already established in academic research (Section~\ref{academic_research}) and first forays can be seen in industrial adoption (Section~\ref{history:industry}).

\subsection{Towards Autonomous Hacking}

We see two research directions leading towards autonomous hacking. The first paper using LLMs for penetration-testing already used autonomous LLM agents~\cite{happe2023getting}. In addition, with increasing delegated task complexity, interactive LLM-systems resemble autonomous hacking systems. When the task becomes ``hack system x'', both approaches converge.

Capability evaluations show that LLMs contain sufficient penetration-testing capabilities to successfully exploit systems, but their reliability is lacking, i.e., the same LLM-driven prototype will find different attack chains within the same testbed during multiple runs (Section~\ref{academic_research}).

Autonomous systems often try to minimize the amount of keeping humans in the loop for efficiency reasons, increasing safety concerns (Section~\ref{concern:safety}) when deploying autonomous prototypes.

We find it concerning that malicious actors already have started to investigate using LLMs for autonomous penetration-testing and influence-operations (Section~\ref{industrial:blackhats}), increasing the need for security tooling with which defenders can test and improve their defenses.

\section{Obstacles to Overcome}
\label{obstacles}

We highlight obstacles that prevent further adaption of LLMs for penetration-testing. We focus on autonomous use as this includes obstacles relevant to vibe-hacking too.

\subsection{Model Features and Stability}
\label{concern:features}

\subsubsection{Impact of Feature Selection}
Agentic AI (Section~\ref{background}) is highly dependent on LLM features such as function-calling and structured-output, limiting model selection to LLMs supporting these features. Empirical research~\cite{li2025thinking} has shown that model support for these features is not homogeneous and, even if features are supported, using these features can impact the overall quality of LLM responses. During our research using LLMs-as-Judges, we saw that switching to structured-output changed the LLM judge's result.

\subsubsection{Minuscule Changes Impact Outcomes.}
\label{concern:minuscle_changes}

LLMs can be unstable and exhibit chaotic behavior: minimal changes to prompts or switching model versions can substantially impact created trajectories. Capability differences between model families are expected, but there can be unexpected differences between versions of the same model family, e.g., there is an ongoing discussion if OpenAI's o1-preview model's capabilities were significantly reduced compared to the final o1 model~\cite{nerfed1,nerfed2}. Obviously, prompt engineering has a large impact upon the LLM's results and their consistency~\cite{wang2024prompt}. Unobvious, formatting changes orthogonal to the prompt's content impact results: He et al.~\cite{he2024doespromptformattingimpact} investigated the impact of using different formats (such as  plain text, Markdown, JSON, and YAML) for providing context information and detected a variance of 40\% when using OpenAI GPT-3.5-turbo. Another problem is that even when using deterministic settings, an LLM's output can still be nondeterministic~\cite{atil2025nondeterminismdeterministicllmsettings}.

These instabilities are problematic as seemingly unrelated prompt adaptions occurring during empirical experiments can potentially impact and taint the experiment's measurements.

\subsection{Safety and Security Concerns}
\label{concern:safety}

\subsubsection{Safety}

As LLMs are autonomously interacting with their environment in potentially destructive ways, safety is of the highest concern. Typically, prototypes embed safety instructions into their prompts, e.g., limit valid targets that the LLM prototype is allowed to attack. If LLMs do not heed those safety instructions, outcomes can be catastrophic. Concurrent research investigates potential catastrophic fallout when LLMs are used in National Security Applications~\cite{caballero2025large} or CBRN~\cite{xu2025nucleardeployedanalyzingcatastrophic} domains.\footnote{Chemical, Biological, Radiological and Nuclear (CBRN)} Safety instructions are also employed during penetration-testing. Happe et al.~\cite{happe2025llmshackenterprisenetworks} describe while they instructed LLMs to only target their lab network range, some of their evaluated LLMs ignored those instructions and attacked systems explicitly forbidden form being targeted.

\subsubsection{Alignment Concerns}

Happe et al.~\cite{happe2025llmshackenterprisenetworks} highlight another case in which the LLM diverged from the user's original intent by discarding the assigned task and starting to solve an unrelated security task, thus breaking the model's alignment with the user's goals. Similar problems can occur unintentionally, e.g., through changes in the model's deployment infrastructure~\cite{ghandeharioun2024s}, or can be maliciously precipitated~\cite{lynch2025agentic}.

\subsubsection{Security}

Safety and security are intertwined. While we are investigating the offensive use of LLMs, offensive prototypes can also become the target of adversaries, e.g., became victims to active or forward defenses. Offensive AI agents are high-value targets for attackers as they sit at the intersection of private data, untrusted content, powerful actions, and external communication~\cite{trifecta}. An adversary, that is able to take over an offensive AI agent, e.g., through using a prompt injection, gains powerful means of attacking the agent's owner or an unrelated third-party system, further complicating attribution.

\subsection{Costs and Efficiency Concerns}
\label{concern:costs}

\subsubsection{Monetary Costs} Running a LLM depends upon a costly runtime environment: either they run on expensive local AI accelerators or are hosted within on-demand clouds occurring a per-minute cost. Especially newer reasoning models can incur unexpected costs when using their maker's cloud offerings: spending thousands of US\$ for running a single prototype for few hours is not unheard of; using a non-reasoning LLM with the same prototype can cost 1--2 orders of magnitude less.

\subsubsection{Ecological Costs} Concerns about non-monetary cost are becoming more prominent~\cite{ecologic_costs}. Energy-usage of LLMs imposes a ecological burden that needs to be answered for by their utility. Analysis of the energy and water usage of different model families indicates that reasoning models, such as OpenAI o3 or DeepSeek-R1, consume 70 times the energy compared to a small LLM such as OpenAI GPT-4.1-nano, making conscious model selection an important goal for sustainability~\cite{jegham2025hungryaibenchmarkingenergy}.

\subsubsection{Effectiveness of Using LLMs}
\label{concern:shellscript}

Given the mentioned monetary and ecological costs, usage of LLMs should show cost-effectiveness. The common internet saying of ``\textit{go away or I will replace you with a small shell script}'' encapsulates the difference between the two extremes: writing a shell script can be tedious and resource-intensive, but operating it is light on resources. Creating a LLM-driven hacking prototype is comparatively cheap, but running it is resource-intensive. Coming back to Sommer and Paxson~\cite{sommer2010outside}, machine-learning might not be an end in itself, but rather an under-appreciated means to an end, used to illuminating the problem space.

\subsection{Privacy and Digital Sovereignty Concerns}
\label{concern:privacy}

LLMs do not exist in emptiness but within the real-world with its geo-political tensions.

\subsubsection{Privacy}

When LLM-driven penetration-testing prototypes interact with enterprise networks, there's a high chance of them ingesting sensitive data. If using cloud-provided LLMs, this sensitive data is transported to the cloud of the respective LLM provider---a potential violation of digital sovereignty. If an LLM-driven prototype is able to detect and exploit a vulnerability, knowledge of this vulnerability is also transported to the LLM cloud; this means that knowledge of an exploitable vulnerability is thus stored outside of the security perimeter and control of the affected company---both a security and privacy risk.

\subsubsection{Digital Sovereignty}

LLMs are currently only provided by a limited amount of countries, making LLM users dependent upon those countries. 

Another problem is the opaque nature of LLMs. Due to their missing explainability, both closed- and open-weight models can contain backdoors that can result in adversary-planted behavior when triggered by a predefined interaction with the environment. Kutosov et al.~\cite{kutasovshade} tasked frontier LLMs to compete a main task while they should also covertly execute a side-task, emulating industrial espionage, sabotage, and insider threats. Complex models such as Claude 3.7 Sonnet and Gemini 2.5 Pro performed ``best'', resulting in 27\% and 15\% successful covert task execution.

\subsection{Accountability Concerns}
\label{concern:accountability}

\subsubsection{Who is accountable if a LLM makes a faulty decision?}

A famous quote from a 1970's IBM manual has become surprisingly relevant again: ``\textit{A computer can never be accountable therefore a computer must never make a management decision}''~\cite{ibm}. There has been legal precedent of a company being made liable for advice given by a LLM-powered chat bot~\cite{bbc_liable}.

Given the potential catastrophic impact of a security LLM-prototype interacting with their target environment, who is liable for direct and collateral damages? Imagine, that a company's operation is disrupted due to a LLM-operated penetration-test that executed a destructive command unrelated to the penetration-testing task at hand.  Who is liable for damages? What if this happens not to a ``normal'' company but to a power plant?

\subsubsection{Overlap with Explainability}

Regulatory and legal requirements often imply explainability. The European Union's AI Act mandates that organizations operating high-risk AI systems must provide meaningful explanations of AI decisions to individuals. In the United States of America, \textit{the FDA's proposed framework for medical devices empathizes the importance of explainable AI for patient safety and clinical decision making}~\cite{regulatory_explainability}.

\subsection{Capability vs. Reliability}

\subsubsection{Missing Reliability}
\label{concern:reliability}

As mentioned in Sections~\ref{history:breaching} and \ref{industrial:blackhats}, experiments indicate that LLMs exhibit capabilities for penetration-testing but lack reliability, i.e., multiple runs of the same prototype against the same testbed yield different vulnerabilities. The obvious solution of repeatedly calling the prototype is problematic due to the higher time- and resource usage (Section~\ref{concern:costs}). In addition, if a security-test should be covert, repeatedly calling the prototype might trigger detection. Better approaches to raise consistency and reliability are needed.

\subsubsection{The Problem with Capability Evaluations}
\label{concern:evaluation}

Capability Evaluations are used to measure LLM's abilities throughout diverse fields, typically using testbeds and benchmarks. Benchmark test-cases have multiple desired properties, e.g., atomicity of test-cases and reproducibility, that can conflict with real-world use-cases which are often ``messy'' and not reproducible~\cite{happe2025benchmarkingpracticesllmdrivenoffensive}. There are concerns~\cite{sommer2010outside,lukošiūtė2025llmcyberevaluationsdont} that synthetic test-beds often do not measure real-world impact and thus not provide meaningful guidance for LLM development.

\subsection{Concerns About Ethics}
\label{concern:ethics}

While usage of LLMs might be technically feasible, the question arises if it is ethical, or wise, to advance this research. Machine-learning ethics is a diverse field; we only cover the subset of using LLMs for penetration-testing purposes and refer to existing publications for areas not covered in this publication, such as bias~\cite{taubenfeld2024systematic,wan2023kelly,dai2024bias} or potential problems with training data~\cite{maini2024llm,draper2023potential,cooper2025extractingmemorizedpiecescopyrighted}.

\subsubsection{Democratizing Access to Penetration-Testing}

Mirsky et al.~\cite{mirsky2023threat} noted that \textbf{AI is a Double-Edged Sword}. Security researchers with noble intentions see the potential of AI democratizing access to security testing, i.e., providing access to security testing to parties that currently lack means of testing to improve their security posture. Examples typically given include small- and medium businesses (SMEs), non-governmental, and non-profit organizations (NGOs and NPOs). As shown in our review of blackhat activities (Section~\ref{industrial:blackhats}), malicious actors also see the potential of AI, although with less benign intentions. This problematic situation results from the inherent dual-purpose nature of security tooling~\cite{lupovici2021dual}. Recent research into how security researcher express their ethical considerations indicates that they are aware of this ethical dilemma~\cite{happe2025ethicsusingllmsoffensive}.

\subsubsection{Impact Upon Workforce}

In economics, Schumpeter's \textit{Creative Destruction} describes a process in which new innovations replace and make obsolete older innovations~\cite{diamond2006schumpeter}. The impact of LLMs, especially their potential for automating tasks, on the workforce is currently subject of academic discussions~\cite{koumpan12024revolutionizing, eloundou2024gpts, wang2023large,herrera2024ai,resh2025complementarityaugmentationsubstitutivityimpact,shao2025futureworkaiagents} with the hopeful expectation that LLMs will be beneficial for novice and lower-skill workers~\cite{brynjolfsson2025generative}.

Experience indicates that successful use of autonomous LLM-agents depends on oversight by highly-skilled human workers with domain knowledge~\cite{oversight}. Recent research shows that (premature) use of LLMs during education, e.g., through delegating information-retrieval tasks, has negative impact on brain neural connectivity, leading to measurable negative impact on learning skills~\cite{kosmyna2025brainchatgptaccumulationcognitive}. Do we bite the hand that feeds us?

% Just how deep do you believe?
% Will you bite the hand that feeds?
% Will you chew until it bleeds?
% Can you get up off your knees?
% Are you brave enough to see?
% Do you wanna change it?
%
% NiN --- The Hand that Feeds

\section{The Way Forward?}
\label{conlusion}

Academic (Section~\ref{academic_research}) and industrial (Section~\ref{history:industry}) uptake of LLMs suggest that they will play an increasing role for cybersecurity in the near future. As Section~\ref{status_quo} showed, vibe-hacking is already here, and real-world use of LLMs for autonomous hacking is currently investigated by both white- and blackhats. Based on the issues mentioned in Section~\ref{obstacles}, we now turn to potential remediations and research opportunities to enable and ease adaption of LLMs for offensive security tasks.

\subsection{Costs and Features}

While LLM-based experiments can impose substantial costs (Section~\ref{concern:costs}), the overall trajectory indicates continuously decreasing costs per token. For example, during January 2025, one million output/reasoning tokens using OpenAI's o1 model would cost \$60. Five months later, using the new o3 model, the costs would have been reduced from \$60 to \$8 per million output/reasoning tokens.

Newer model releases typically also improve their feature support, e.g., function-calling and structured-output. Currently, reasoning models seem to struggle with function calling~\cite{berkeley-function-calling-leaderboard}. While this problem is something that researchers should be aware of, this situation should resolve itself quickly.

While costs per token decreases, overall volume of token consumption is rising and thus impacts global energy use and ecology. To reduce this impact, careful selection of models is essential, e.g., only using reasoning models if their capabilities are needed. Unfortunately, one of the most resource-intensive areas of using LLMs, image- and video-creation~\cite{llm_ecology}, is used by blackhats for influence operations (Section~\ref{industrial:blackhats})---we assume that these malicious operators are not that ecology-conscious. Ultimately, the decision about the ecological impact of LLMs if for each individual, and societies at large, to decide (Section~\ref{society_uptake}).

\subsection{The Need for Better Safeguards}

LLMs become further integrated into security workflows and thus gain more potential to interact with their environment, making their security and safety parameter (Section~\ref{concern:safety}). If security or safety incidents occur, Accountability (Section~\ref{concern:accountability}) becomes important.

Google identified multiple features needed for operating secure LLM-based systems~\cite{1034542}. Prominent features were, e.g., protection from malicious inputs, limitation of agents' interaction capabilities, and clear human oversight being enforced.

Prototypes can use LlamaGuard~\cite{inan2023llama} to analyze user input for unsafe content, e.g., if a user might follow criminal intentions. The similarly named Llama PromptGuard~\cite{prompt_guard} can be used to detect malicious inputs used for prompt-injection or jailbreak attacks. Google CaMeL~\cite{debenedetti2025defeating} applies \textit{Control Flow Integrity}~\cite{abadi2009control}, a traditional defensive mechanism, to LLM prompting and uses it to separate data from control flow. Meta's CodeShield~\cite{chennabasappa2025llamafirewall} uses source-code analysis to prevent malicious program generation through agentic AI. Agent AlignmentChecks~\cite{chennabasappa2025llamafirewall} continuously compares a LLM's reasoning trace with the user's stated goal to detect misalignment.

Still, using agentic AI for penetration-testing imposes the problem that we cannot differentiate ethical hacking (whitehats) from unethical hacking (blackhats) as the utilized mechanisms and techniques are identical. In the case of red-teaming, there is not even a difference in the target's awareness of being attacked. Ultimately, the only difference is the potential impact of a penetration-testing campaign on the target, but as this is orthogonal to the penetration-testing operations that were performed before,  this cannot be used to differentiate ethical from unethical behavior.

Please note that concerns for safety and security should not be abused as reason to put ethical penetration-testing prototypes inside walled-gardens to prepare later commercialization.

\subsection{Capabilities and Reliability}

As seen in Section~\ref{concern:reliability}, there is currently a lack of reliability and reproducibility when using LLMs for penetration-testing. Improvements can be categorized in single-agent and multi-agent solutions. The former try to improve reliability within a single agent's trajectory, while the latter combine multiple agents to increase reliability while typically imposing higher token costs due to more agents being run.

Single-LLM solutions currently explore the impact of self-discussion or self-challenging~\cite{zhou2025self}, investigate better state management~\cite{laban2025llms}, or implement auto-correction mechanisms~\cite{happe2025llmshackenterprisenetworks}. Multi-LLM solutions typically combine output of multiple agents using techniques such as LLM-as-judge~\cite{gu2024survey} to integrate the results of singular agents. As Anthropic notes~\cite{anthrophic_multiagent}, multi-agent systems improve the actin-space search breadth but incur substantial higher token costs (they note that their multi-agent system consumes $3.75$ times the tokens of a similar single-agent system) introducing an overlap with monetary and economic cost concerns (Section~\ref{concern:costs}).

If a multi-agent system uses multiple different LLMs, concerns about LLM stability in face of minuscule changes altering the trajectory substantially (Section~\ref{concern:minuscle_changes} should be alleviated as each individual model should have a reduced overall impact on the system's result.

Regardless of the chosen approach, more empirical data about model behavior is needed. Improving the explainability of models would make analysis of models' decisions more impactful and also more efficient.

\subsection{Decision Time for Individuals and Society}
\label{society_uptake}

While we try to propose technological improvements, some of the imposed problems of LLMs are for society to decide, i.e., are a socio-economical problem, not a technological one. We need consensus on how to handle privacy, ecological impact, ethical issues, accountability, and digital sovereignty concerns (Section~\ref{concern:ethics}). While technology solutions for parts of the problems exists, e.g., running small language models locally to keep data private, individuals will often have to ultimately decide if they value, e.g., privacy, over LLM-provided features.

Andy Masley compared resource-consumption of using LLMs with other lifestyle-decisions~\cite{ecologic_costs} and calculated the equivalent of ``going vegan'' with $400.000$ chatGPT text queries a year, concluding with that his becoming vegan will offset his chatGPT use. When measuring the average query counts of the two best-performing LLMs for hacking enterprise networks~\cite{happe2025llmshackenterprisenetworks}, we arrive at $165.75$ queries per hours, or $.145$ million queries a year if the autonomous penetration-testing prototype is running 24/7. Given their preliminary results, this could prevent dozens or hundreds of ransomware incidents saving resource-intesive recovery costs. How would this balance out?

\section{Conclusion}

Using LLMs for offensive security is evolving at a fast pace. Vibe-hacking has already been established; autonomous penetration-testing is currently investigated within Academia while both black- and whitehat hackers are deploying first systems in production. Given the scarcity of professional penetration-testers, a reduction in interest is hard to believe in.

Due to the high-risk environment, keeping humans in the loop is essential for production environments' safety. We hope that the proliferation of using LLMs during in or during education will not reduce human capabilities~\cite{kosmyna2025brainchatgptaccumulationcognitive} needed for performing this oversight.

Security tooling always had an dual-edged nature, especially if it is deployable in an autonomous manner as this reduces the skills needed to perform security audits. We have reports of initial adversarial use (Section~\ref{industrial:blackhats}), we are already within a weapons race. Fittingly, Vernor Vinge~\cite{vinge2007rainbows} used the Red Queen's paradox to illustrate the struggle between encouraging technological advancement and protecting the world if technology is abused. While originally this was written as part of a science fiction story-line, we are living it out right now.

\bibliographystyle{plainnat}
\bibliography{bibliography}

\end{document}